\documentclass[twocolumn,nofootinbib,a4paper,
preprintnumbers,prl]{revtex4-1}

\usepackage[english]{babel}
\usepackage[T1]{fontenc}
\usepackage{graphicx}
\usepackage{amsmath}
\usepackage{amssymb}
\usepackage{amsfonts}
\usepackage{latexsym}
\usepackage{mathrsfs}
\usepackage{color}
\usepackage{slashed}
\usepackage{feynmp}
\usepackage{pstricks,pst-text}
\usepackage{epsfig}
\usepackage{dsfont}

\begin{document}

\title{Effective dynamics of a tracer particle in a dense homogeneous quantum gas}

\author{M. Jeblick}

\author{D. Mitrouskas}
\email[\,]{dmitrous@math.lmu.de}

\author{S. Petrat}

\author{P. Pickl}

\affiliation{%
Mathematisches Institut, Ludwig-Maximilians-Universit\"at, Theresienstr.\ 39, 80333 M\"unchen, Germany}

\begin{abstract}
We investigate the mean field regime of the dynamics of a tracer particle in a homogenous quantum gas. For a bosonic gas, we show that this regime is constrained by the well known requirement of an appropriate mean field scaling of the interaction. For fermions, however, we find an important qualitative difference. Not only are fermions much more homogeneously distributed than bosons but also deviations from the mean are due only to fast degrees of freedom in the gas. This observation leads to an explanation of why a tracer particle behaves freely in the dense homogeneous fermion gas despite of a non-scaled interaction, i.e., despite of non-vanishing statistical fluctuations. Finally, we indicate how the gained insight can be rigorously justified.
\end{abstract}

\maketitle



The time dependent mean field method is a widely used approach for describing the effective dynamics of many body systems. Within this method one approximates the complex microscopic interaction among many particles by an average external potential (the mean field), thus reducing the description to an effective one body problem. Well known examples of such mean field descriptions for different microscopic models in quantum mechanics are the Hartree equation for bosons \cite{hepp:1974,spohn:1980,erdoes:2001,froehlich:2009,rodnianski:2009,pickl:2011method} as well as the Hartree-Fock equation for fermions \cite{bardos:2003,erdoes:2004,froehlich:2011,benedikter:2013}.\\
The question whether such an effective description is accurate or not depends on the statistical fluctuations of the microscopic potential around its average value. More precisely, fluctuations need to be suppressed in such a way that the actual potential $V=\sum v$ felt by a single particle due to all the other particles (whose wave function at a given time $t$ we denote by $\phi_t$) equals its average value $\mathds{E}_{\phi_t}[V]$ in an appropriate thermodynamic (TD) or many particle limit with very high probability:
\begin{align}
		\label{Ginzburg}
\mathds{P}_{\phi_t}\left (  V  \approx   \mathds{E}_{\phi_t}[V] \right) \approx 1.
\end{align}
Heuristically, this criterion provides a sufficient condition for the replacement of the microscopic potential by its average value. It is therefore often used as defining the so called mean field regime, i.e., the range of applicability of the mean field description. In general, it requires a potential $v$ whose strength becomes weaker with increasing number of particles $N$ in the gas, as, e.g., in the case of the Hartree equation: $v = \mathcal O (N^{-1})$. The scaling of the microscopic potential diminishes the fluctuations so that they become negligible in the thermodynamic limit which implies eq.(\ref{Ginzburg}). Given an appropriate product structure of the initial state of the gas, the dynamics of any of the particles is then typically determined by the appropriate mean field potential 
or, in other words, the mean field description provides a good approximation of the microscopic dynamics. In contrast, for strong microscopic interactions, i.e., $v=\mathcal O(1)$, eq.\eqref{Ginzburg} is usually not fulfilled and from this it is often concluded that the mean field description breaks down due to the non-vanishing impact of fluctuations.\\

This brings us to the objective of this letter. Our claim is that there are interesting situations in which the condition given by eq.(\ref{Ginzburg}) and thus a restricting scaling assumption is not necessary for the accuracy of the mean field description. The particular situation we investigate here is the motion of a tracer particle in a homogeneous and dense quantum gas for which the effective description corresponds simply to the free time evolution. By a careful analysis of the statistical fluctuations, we identify an important difference between bosonic and fermionic quantum gases that, to our knowledge, has not appeared in the literature before. Our demonstration indicates that the applicability of the mean field method for homogeneously distributed bosons is given, as expected, only under the assumption of an appropriate mean field scaling which ensures that eq.\eqref{Ginzburg} holds. In case of fermions, on the contrary, it turns out that the condition given by eq.(\ref{Ginzburg}) is actually not necessary for the mean field description to be a good approximation of the microscopic dynamics. We give a detailed explanation of why for a dense homogeneous fermion gas the mean field method is accurate without any restricting scaling assumption. This finding constitutes our main result. For the sake of clarity, the argument is demonstrated for a tracer particle in a non-interacting quantum gas in one dimension. For this model we give a short summary of how our main result can be rigorously justified. However, we emphasize that the physical argument we present is very general, so that it applies to interacting gases in higher dimensions as well.

\vspace{-12pt}

\section{\label{Model}Microscopic model}
 
The combined system of tracer particle and bosonic resp.\ fermionic $N$ body system (quantum gas) is defined on a one dimensional torus $\mathbb{T}\subset\mathbb{R}$ of length $L$, i.e., in a box with periodic boundary conditions, and described by a square integrable wave function $\Phi_t^{\pm} \in L^2(\mathbb{T},dy) \otimes L^2_{\pm}(\mathbb{T}^N,dX)$. The coordinates of the gas particles are denoted by $X=(x_{-N/2},...,x_{N/2})$ while the superscripts $+$ and $-$ refer to the bosonic and fermionic case, respectively. The microscopic time evolution is determined by the Schr\"odinger equation 
\begin{align}
		\label{SchroedingerGl.}
		i\dot{\Phi}^{\pm}_t=H \Phi_t^{\pm}
\end{align}
(for ease of notation we set $\hbar=1=2m$) with the $N+1$ particle Hamiltonian
\begin{align}
			\label{Hamiltonian}
			H=-\Delta_y-\sum_{\vert j\vert \le N/2}\Delta_{x_j}+\sum_{\vert j \vert \le N/2}v(x_j-y).
\end{align}
The pairwise interaction between tracer particle and gas particles is modeled, for simplicity, by a rectangular function of height $v_0=\mathcal O (1)$ and width $l_0$ (we emphasize again that the chosen potential is not scaled as it is usually the case in the derivation of mean field equations). The interaction term in eq.(\ref{Hamiltonian}) is abbreviated by $V=\sum v$.\\
We are interested in initial conditions with a distinct product structure (we choose $t_0=0$): 
\begin{align}
			\label{Produkt-Struktur}			
			\Phi_{0}^{\pm}=\chi_{0}\cdot \phi^{\pm}_{N_{0}},
\end{align}
$\chi_{0}\in L^2(\mathbb{T},dy)$ and ${\phi^{\pm}_{N_{0}}}\in L^2_{\pm}(\mathbb{T}^N,dX)$ being appropriate initial wave functions for the tracer particle and for the gas. In the case of bosons, we consider two different possible initial states. On the one hand, a so called Hartree state or condensate, i.e., a product of one particle wave functions:
\begin{align}
			\label{Kondensat}
		    {\phi^{+_1}_{N_{0}}}(X)=\prod_{\vert j \vert \le N/2} {\varphi_{p}}_0(x_j),
\end{align}
where ${\varphi_{p}}_0$ is a normalized solution of the free one particle Schr\"odinger equation:
\begin{align}
			\label{FreeSol}
			{\varphi_{p}}_0(x)=\frac{1}{\sqrt{L}}e^{i p\cdot x}
\end{align}
with arbitrary value of $p$ in the spectrum of allowed momenta: $p_j=2\pi\cdot j/L$, $j\in\mathbb{Z}$.
On the other hand, we consider the symmetric product of free one particle solutions (\ref{FreeSol}), denoted by ${\varphi_j}_0$, which occupy all possible momenta $p_j$ below the Fermi momentum $p_{N/2}=\pi\cdot \rho$ (we set $\rho=\frac{N}{L}$):
\begin{align}
			\label{Bosonen}
		   {\phi^{+_2}_{N_{0}}}(X)=\prod_{\vert j \vert \le N/2}^{sym.} {\varphi_j}_{0}(x_j).
\end{align}
The fermionic gas is initially assumed to be the non-interacting ground state of the $N$ body system, i.e., a slater determinant of plane waves ${\varphi_j}_0$, $\vert j\vert \le N/2$:
\begin{align}
			\label{Fermionen}
		   {\phi^-_{N_{0}}}(X)= \prod_{\vert j \vert \le N/2}^{asym.} {\varphi_j}_{0}(x_j).
\end{align}

\section{Mean Field Description}

There are two questions which arise with regard to the accuracy of the mean field description in the addressed situation. On the one hand, whether the initial homogeneity of the gas is disturbed by the presence of the tracer particle and, on the other, whether the dynamics of the tracer particle is determined by an effective one body equation.\\
The fermionic ground state of a dense gas is very robust against the external potential caused by the tracer particle. The reason for this is that gas particles with energy in the range of $v_0$ can hardly be excited due to Pauli's principle whereas particles occupying states close to the Fermi energy remain almost undisturbed by the external potential $v_0 \ll E_F= (\pi\cdot \rho)^2$. It can be shown that the time evolution of the fermionic gas (\ref{Fermionen}) decouples completely from the external potential $v$ in the limit of very high densities so that the gas evolves effectively freely:
\begin{align}
				\label{MFFer}
				\phi_{N_t}^{-mf}=e^{-i \sum_j \Delta_j  t} \phi_{N_{0}}^{-} = \phi_{N_{t}}^{- f}.
\end{align}
The superscripts $mf$ and $f$ refer to mean field and free time evolution, respectively. This does not hold for the other two initial states (\ref{Kondensat},\ref{Bosonen}) though. Since $v_0$ is assumed to be $\mathcal{O}(1)$, the bosonic gas is disturbed by the presence of the tracer particle which makes the microscopic dynamics more complicated. But it turns out that this complication is not important with regard to the analysis of the validity of the mean field method. Our demonstration shows that the tracer particle does not behave according to the mean field description even if one neglects this complication. We therefore disregard it from now on and assume the bosonic gas to evolve independently as well:\footnote{Note that, once an appropriate mean field scaling of the potential $v$, e.g., $v=\mathcal O (N^{-1})$, is assumed, a boson gas of many particles does no longer feel the disturbance due to one single tracer particle.}
\begin{align}
			\label{MFBos}
			\phi_{N_t}^{+mf}=e^{-i \sum_j \Delta_j  t} \phi_{N_{0}}^{+} = \phi_{N_{t}}^{+f}.
\end{align}
The average potential produced by the gas particles determines the mean field description of the tracer particle:
\begin{align}
	 \mathbb{E}_{{\phi_{N_t}^{\pm f}}}[V](y) = \left({\phi_{N_t}^{\pm f}}\Big| \sum v(\cdot-y) {\phi_{N_t}^{\pm f}} \right) =v_0\cdot \rho l_0,
\end{align}
where we denote by $(\cdot |\cdot )$ the scalar product in $L^2(\mathds{T}^N,dX)$. Since the average potential is spatially as well as temporally constant, the mean field dynamics equals the free time evolution up to a constant phase:
\begin{align}
		\label{mfdesc}
		\chi_t^{mf}=e^{-i (v_0\cdot \rho l_0)t} \chi_{t}^{f}.
\end{align}
There is a helpful intuitive picture behind this equation. The tracer particle is surrounded by a gas consisting of $N$ freely moving particles. The mean fraction of particles $\rho l_0$ in the interacting neighborhood of the tracer particle is constant. This is why the forces from left and right cancel each other on average. The actual number of gas particles, however, might randomly deviate from its average value, which, in turn, generates a gradient in the potential and consequently a non-vanishing force on the tracer particle. According to this picture it is the deviations from the potential caused by random fluctuations around the average number of particles which determines whether the predictions of the mean field method (\ref{mfdesc}) are accurate or not.

\vspace{-8pt}

\section{Fluctuations}

That eq.\eqref{Ginzburg} states a sufficient condition for the validity of the mean field description in our model can be directly inferred from the comparison of the microscopic dynamics with the mean field time evolution:
\begin{align}
			\label{Comp}
			\left \vert\left \vert  e^{-iH t} \Phi_{0}^{\pm}  - \chi_{t}^{mf}  \cdot \phi^{\pm f}_{N_{t}} \right \vert \right \vert  \le t \cdot \sqrt{\text{Var}_{{\phi_{N_t}^{\pm f}}} [V]}.
\end{align}
The vanishing of the right-hand side depends on the magnitude of fluctuations of the potential. The latter determines the likeliness of a deviation of the potential from its average value and can therefore be understood as a measure of the strength of the randomly acting forces in the gas. The above inequality thus expresses the same condition as eq.(\ref{Ginzburg}): the mean field description provides a good approximation if all random forces in the gas disappear. As is shown in the following subsection, the fluctuations of the potential do not vanish for neither of the three addressed cases (\ref{Kondensat},\ref{Bosonen},\ref{Fermionen}) which is due to the fact that $v=\mathcal{O}(1)$.\\
In order to understand whether eq.\eqref{Ginzburg} is also a necessary condition for the applicability of the mean field method, it is helpful to analyze another property of the statistical fluctuations, namely the time scales on which they typically appear. In the heuristic picture, those time scales correspond to the duration of the randomly acting forces on the tracer particle. The latter are, in turn, determined by the momenta of those particles causing the fluctuations around the mean. A force produced by deviations due to slow particles acts, e.g., longer than a force caused by fluctuations due to fast ones. We therefore investigate which of the gas particles actually tend to deviate. The combination of both properties, magnitude and typical duration, offers a heuristic estimate not only of the strength of the forces but of the magnitude of the randomly transferred momentum (\textit{force $\times$ time $=$ momentum}) to the tracer particle which we denote by $\delta p_{y}$. Thus, we suggest the following weaker but physically more relevant condition for the accuracy of the mean field description \eqref{mfdesc} in the addressed situation: $\delta p_y\approx 0$ in the appropriate thermodynamic limit.\\
In case of bosons, this condition turns out to coincide with eq.\eqref{Ginzburg}, which, as is shown below, requires an appropriate mean field scaling. For fermions, however, we show that the random momentum transfer vanishes without any scaling assumption and despite eq.(\ref{Ginzburg}) does not hold.

\vspace{-10pt}

\subsection{\label{Mag}Magnitude of fluctuations}

One can think of $\{v(x_j-y)\}_{\vert j \vert \le N/2}$ as a family of random variables whose sum $V=\sum v$ assumes different moments for different distributions. The latter are determined by the corresponding wave function describing the gas. The condensate (\ref{Kondensat}), e.g., is characterized by its product structure which defines a sequence of independent and identically distributed random variables. The fluctuations behave in this case according to the $\sqrt{N}$-law:
\begin{align}
			\label{FluktuationenB1}
			\text{Var}_{\phi^{+_1}_{N_t}} [V] (y) = v_0^2\cdot \rho l_0.
\end{align}
In the symmetrized boson state (\ref{Bosonen}) the random variables are correlated. Nevertheless, the fluctuations behave in the thermodynamic limit (i.e., for $N\to\infty$, $\rho=const.$) similar to the condensate:
\begin{align}
		\label{FluktuationenB2}
			\lim_{TD} \text{Var}_{{\phi^{+_2}_{N_t}}} [V](y)  = \mathcal O \left( v_0^2\cdot \rho l_0 \right).
\end{align}
The correlations of the antisymmetric fermionic ground state wave function (\ref{Fermionen}) lead to an expected decrease in the magnitude of fluctuations:
\begin{align}
		\label{FluktuationenF}
		\lim_{TD} \text{Var}_{\phi^{-}_{N_t}}[V](y) = \mathcal O \left( v_0^2 \cdot \ln(\rho l_0)\right).
\end{align}
Physically, one can say that the Fermi pressure causes a comparatively much more homogeneous distribution of the gas. 

\vspace{-6pt}

\subsection{\label{Time}Which particles fluctuate?}

In the suggested heuristic picture, the gas consists of freely moving particles with momenta according to the occupied plane waves. If one thinks of a random force as being caused by a deviation of the average particle number due to a particle with momentum $p$, then such a force acts on the tracer particle for a time $t=l_0/\vert p\vert$. This is simply the length of time needed for the particle to pass the range of interaction with the tracer particle. In the condensate, all particles have the same momentum $p$, which is why the typical time scale of the random forces is given by $t=l_0 /\vert p\vert $. For the symmetric and antisymmetric product states, particles occupy momenta between zero and the Fermi momentum: $0\le |p|\le p_F= \pi\cdot \rho$. In order to analyze which of the particles produce the fluctuations in the gas it is useful to rewrite the respective total magnitude  (\ref{FluktuationenB2},\ref{FluktuationenF}) in terms of a sum over all occupied momenta in the gas:
\begin{align}
			\label{TimeScales}
			\text{Var}_{\phi_{N_t}^{\pm}}[V]= \sum_{\vert j \vert \le N/2} \text{var}_{{\phi_N}_t^{\pm}}[V](p_j).
\end{align}
The hereby obtained function $\text{var}[V](p_j)$ can be interpreted as a measure of likeliness of a deviation due to a particle with momentum $p_j$. The explicit expressions for $\text{var}[V](p_j)$ are listed in the appendix in eq.(\ref{Append}). The possible time scales are now given by $t_j=l_0 / \vert p_j\vert$ with $p_j=2\pi\cdot j/L$ and $|j|\le N/2$. The qualitative behavior of the function $\text{var}[V](p_j)$ for different gas densities is shown in Fig.(1)-(3).\vspace{0.5cm}

\includegraphics[width=80mm , height=45mm]{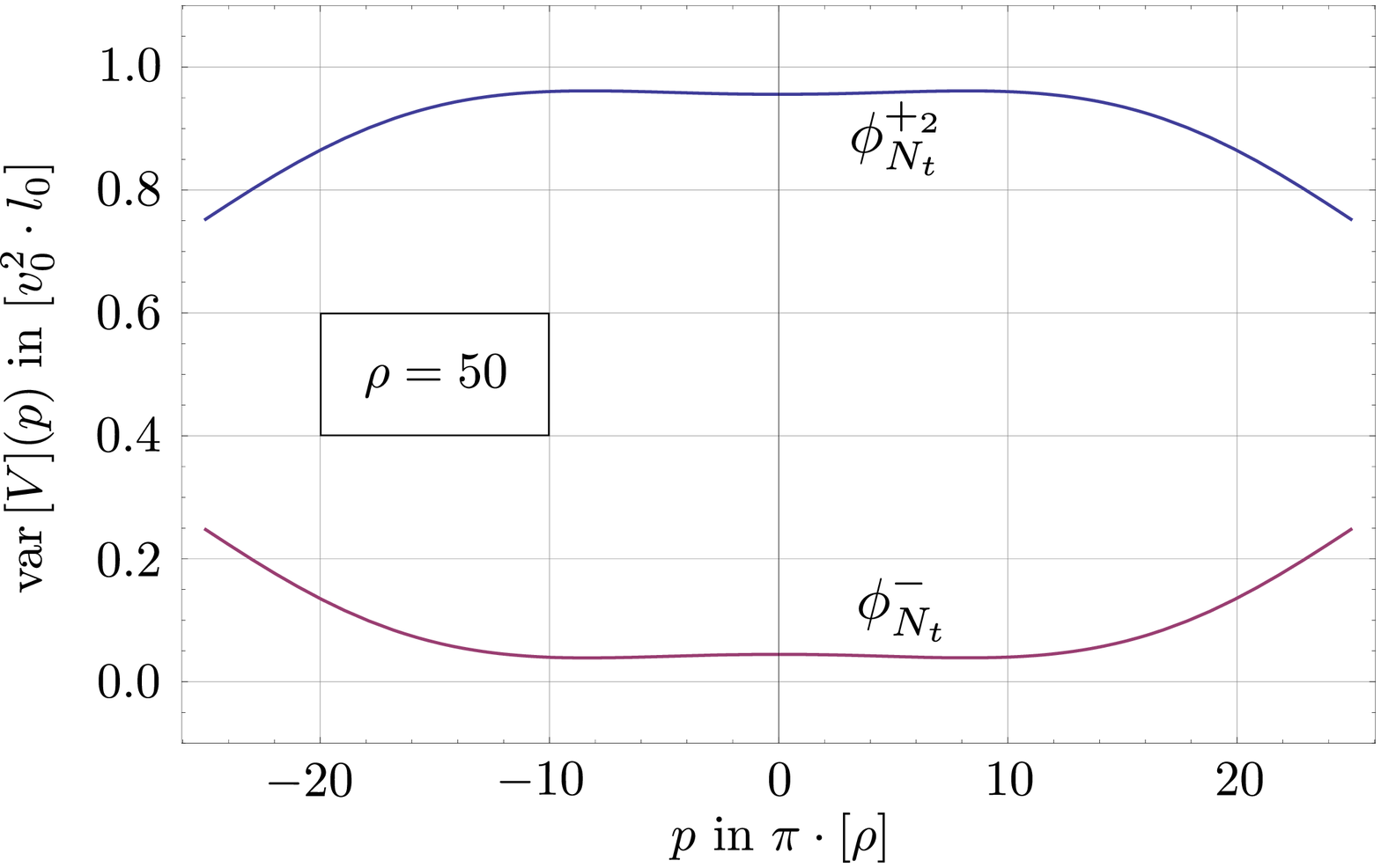}

\vspace{0.5cm}

\includegraphics[width=80mm , height=45mm]{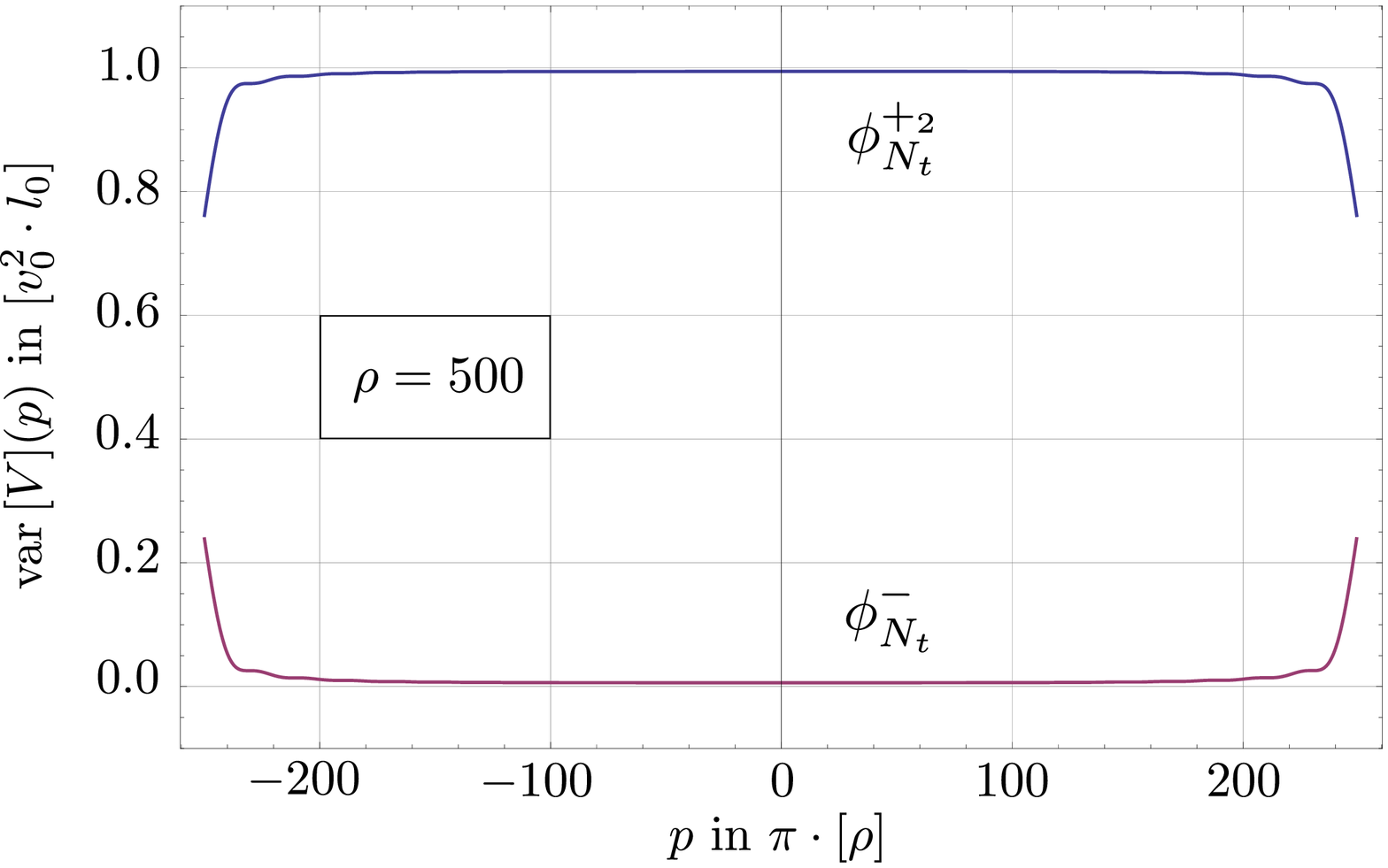}

\vspace{0.5cm}

\includegraphics[width=80mm , height=45mm]{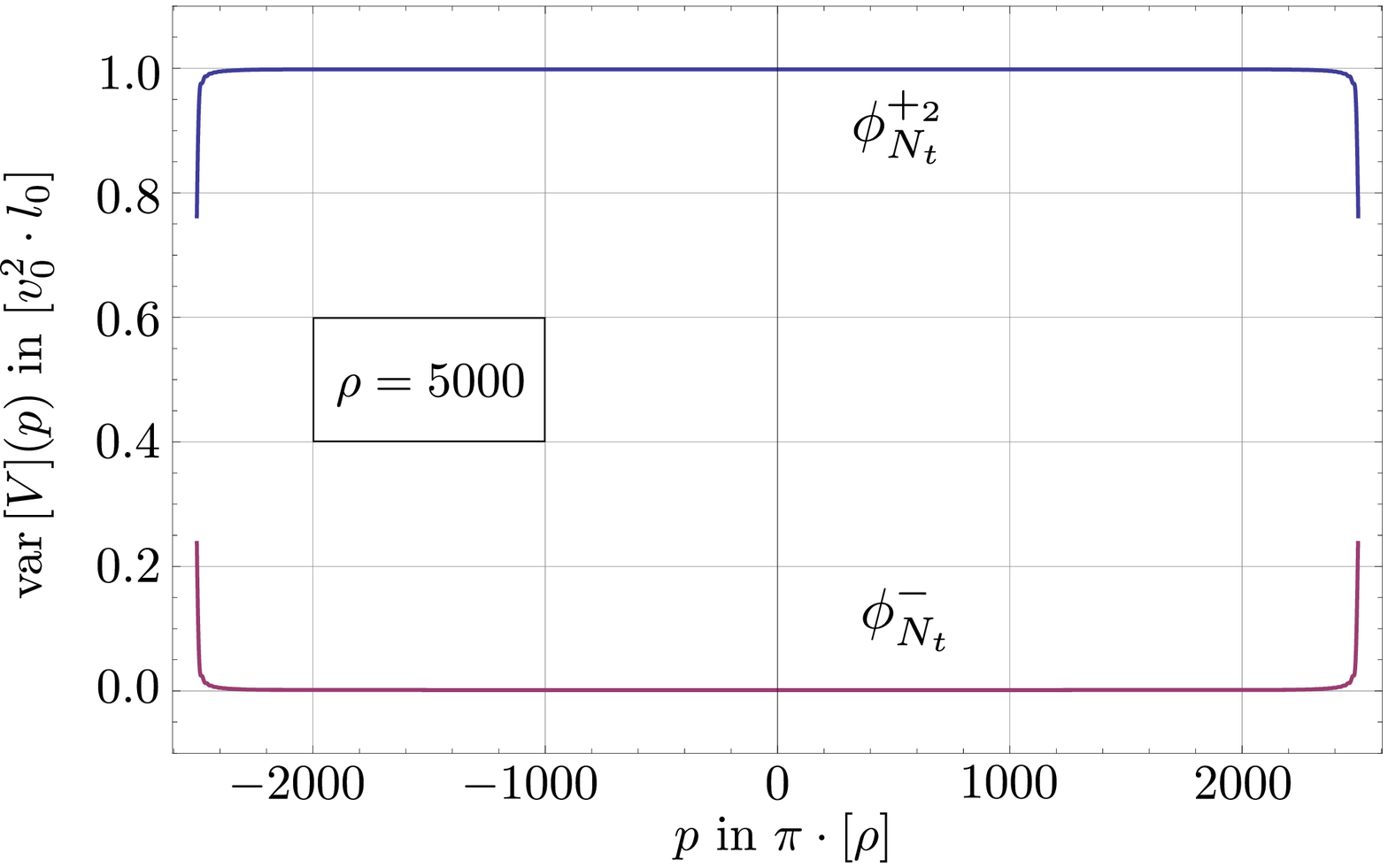}
\begin{center}
\small{Fig.(1)-(3): var[$V$] for $\rho=50$, $500$ and $5000$}
\end{center}

What can be inferred from the above figures? The upper curve indicates that in the bosonic gas almost all occupied momenta contribute an equal amount to the total magnitude of fluctuations (the total magnitude of the fluctuations is according to eq.(\ref{TimeScales}) approximately given by the area under the respective curve). This means that it is equally likely for all particles in a bosonic gas to produce a deviation from the average number of particles in the neighborhood of the tracer particle. This is again very similar to the behavior of the condensate.\\
The lower curve, on the contrary, shows that the probability to fluctuate decreases rapidly for the slow degrees of freedom in a dense fermionic gas. The increasing wings are pushed outwards to the high momenta. It can be further shown that fluctuations due to particles with momenta $\vert p_j\vert \le \pi\cdot \sqrt{\rho}$ vanish completely for high densities:
\begin{align}
	\label{SlowDegrees}
		\lim_{\rho\to \infty} \lim_{TD} \sum_{\{ j: \vert p_j\vert \le \pi\sqrt{\rho} \}} 					\text{var}_{{\phi_N}_t^-}[V](p_j) = 0.
\end{align}
This implies that the actual number of gas particles with slow momenta in the neighborhood of the tracer particle converges to its average value. The deviation of the potential from the mean field is therefore caused only by particles with very high momenta. Random forces appear, so to speak, typically on very fast time scales compared to the bosonic gas.\\

\vspace{-16pt} 

\section{\label{Validity}Validity of mean field description}

\vspace{-5pt}

\subsection{Boson gas}

Our findings strongly suggest the failure of the mean field description (\ref{mfdesc}) as a good approximation of the actual behavior of the tracer particle in the bosonic gas. Neither is eq.\eqref{Ginzburg} fulfilled [see eqs.(\ref{FluktuationenB1}) and (\ref{FluktuationenB2})] nor does the random momentum transfer on the tracer particle vanish. The forces in the condensate act typically on a time scale $t=l_0/\vert p\vert $ with a strength that increases proportionally to the gas density. Consequently, the effective amount of momentum which is randomly transferred to the tracer particle increases also with increasing density. The situation in a gas described by the symmetrized product state is very similar. The strength of the forces increases proportionally to the gas density while they appear equally probable on all possible time scales $t_j=l_0/\vert p_j \vert $, $\vert j \vert \le N/2$. Thus, the mean field description is not accurate for a tracer particle in a bosonic gas because of the dominant influence of the fluctuations.\\
Note that an appropriate mean field scaling of the microscopic interaction, e.g., the replacement of the potential $v$ in our model (\ref{Hamiltonian}) by the scaled potential $v_{\rho}=v/{\rho}$, suppresses the statistical fluctuations:\footnote{The reason for choosing the potential $v_{\rho}$ instead of, e.g., $v_N=v/N$ is that in the latter case the average potential felt by the tracer particle $v_0 \cdot l_0\rho/N$ vanishes in the thermodynamic limit  ($N\to\infty$, $\rho=const.$) whereas it equals $v_0 \cdot l_0$ for $v_{\rho}$.}
\begin{align}
			\text{Var}_{{\phi_N}^{+}_t}[V](y) =  \mathcal O \left(v_0^2\cdot l_0/\rho \right).
\end{align}
They disappear in the limit $\rho\to \infty$ which ensures that eq.(\ref{Ginzburg}) holds. This directly implies also the weaker condition of a vanishing momentum transfer. The mean field method provides a good description of the dynamics of the tracer particle in this microscopic model which follows directly from inequality (\ref{Comp}). We conclude that, similarly to the well known case of the Hartree equation, the mean field regime of the tracer particle in a bosonic gas is constraint by the requirement of an appropriate scaling of the interaction.

\vspace{-8pt}

\subsection{Fermion gas}

According to eq.(\ref{FluktuationenF}) the criterion defined by eq.\eqref{Ginzburg} is not fulfilled for fermions either. However, Fig.(1)-(3) and eq.(\ref{SlowDegrees}) imply that fluctuations vanish for particles with slow momenta $\vert p_j\vert \le \pi\cdot \sqrt{\rho}$. Put differently, the slow particles in the gas are distributed absolutely homogeneously and do therefore not produce any random forces on the tracer particle. The particles with large momenta, on the contrary, fluctuate.  The strength of the corresponding forces increases proportionally to the logarithm of the density, see eq.(\ref{FluktuationenF}). But fast particles interact only for very short times with the tracer particle: $t_j=l_0 / \vert p_j\vert $ with $\pi\cdot \sqrt {\rho} \le \vert p_j \vert \le \pi\cdot \rho$. Thus, the typical time scales on which the random forces appear decrease inversely polynomially in the gas density: $ l_0(\pi \cdot \rho)^{-1} \le t_j\le l_0(\pi \cdot \sqrt{\rho})^{-1}$ $(\ast)$. The argument is now concluded as follows: the tracer particle behaves freely in a very dense and homogeneous fermionic gas (as in eq.(\ref{mfdesc})) because the randomly appearing forces which are caused by the deviations from the constant average potential do not last long enough in order to transfer a significant amount of momentum to the tracer particle. Indeed, 
\begin{align}
			\label{momFer}
			 \lim_{\rho\to \infty} \lim_{TD}	\delta p_y  \lesssim \lim_{\rho\to \infty} \rho^{-\frac{1}{2}}  \ln(\rho l_0) = 0,
\end{align}
where the upper bound of the momentum transfer is given by the product of the total strength $\sim \ln(\rho l_0)$ \eqref{FluktuationenF} and the longest possible duration $\sim \rho^{-\frac{1}{2}}$, see $(\ast)$, of the random forces produced by fluctuating particles in the fermionic gas.

\vspace{-8pt}

\subsection{\label{Proof}Idea of proof}

The heuristic conclusion we arrived at in the previous subsection constitutes our main result. It motivates the following proposition: the one particle reduced density matrix corresponding to the effective description of the tracer particle in the fermionic gas
\begin{align}
				\mu^{\Phi_t^-}_{red}(y,y') := \int_{\mathbb{T}^N}\ \bar \Phi_t^-(y,X)\cdot \Phi_t^-(y',X)dX
\end{align}
converges (in trace norm, see below) in the thermodynamic limit and for $\rho \to \infty$ to the free one particle density matrix $\mu^{\chi^f_t}=\bar \chi_t^f \cdot \chi^f_t$ with initial condition $\mu^{\chi_0}=\bar \chi_0 \cdot \chi_0$. A complete proof of this proposition and additional generalizations will be presented in \cite{jeblick:2014}. In order to understand the idea of the proof, it is sufficient to consider the interaction in first order perturbation theory, i.e., in the so called Born approximation (note that the constant phase $e^{-i(v_0\cdot \rho l_0)t}$ is omitted on the right-hand side):
\begin{align}
			\label{BornApprox}
		\Phi_t^- - \Phi_t^{-mf} \approx \int_{0}^{t} e^{-iH^f(t-s)} (V -\mathds{E}[V]) \Phi_{s}^{-f} ds.
\end{align}  
Furthermore, we assume smoothness and compactness of the potential, i.e., $v\in \mathcal C_0^{\infty}(\mathds R)$ which simplifies the mathematical proof, while leaving the physical argument unchanged. Within the named approximation a straightforward calculation leads to
\begin{align}
		 &\lim_{TD}\left\vert\left\vert \mu_{red}^{\Phi_t^-} - \mu^{\chi_t^f} \right\vert\right\vert_{tr} \le 
\underset{\vert k \vert \ge \frac{\rho}{2}}{\int dk} \underset{\vert p \vert \le \frac{\rho}{2}}{\int dp} 
\frac{\left\vert  \hat v(k-p)\right\vert^2}{(k^2-p^2)^2} \\ 
		 &\ \ \ \ \ \ \ \ \ \times \Bigg\vert\Bigg\vert  \int\limits_{0}^t\ e^{i\Delta_y s }[e^{i(k-p)y}\chi_s^{f}(y)] \cdot \frac{d}{ds} e^{i(k^2-p^2)s}ds\Bigg\vert\Bigg\vert^2_{y}, \nonumber
\end{align}
where $\hat v$ denotes the Fourier transform of the potential. That the right-hand side vanishes in the limit of high densities can be seen by separating the range of integration into two distinct parts: one which closely enfolds the Fermi edge: $\vert k-p\vert \le 1 / \sqrt{\rho}$. This contribution vanishes since it is bounded from above by $(\vert\vert v \vert \vert_1 \cdot t)^2 /\rho$. The second part for which $\vert k-p\vert \ge 1/\sqrt{\rho}$ holds, becomes suppressed in the case of high densities due to the fast oscillating phase: $(k^2-p^2)\ge 2\sqrt{\rho}$. This can be inferred after integrating by parts, applying Stone's theorem and using the sufficiently strong decay properties of $\hat v$. Note that the above estimate reflects exactly the physical argument we gave in the previous subsection and is, indeed, very reminiscent of the heuristic estimate in inequality (\ref{momFer}).

\vspace{-4pt}

\section{Outlook}

Although the argument was demonstrated on behalf of a simple model in one dimension, the gained insight about the physics of a many body fermion system applies to more complicated models as well. From what has been said, one can conclude that the mean field description for a tracer particle in a dense fermion gas is accurate whenever the average density of the fast degrees of freedom in the gas is homogeneous. Interestingly, this situation is often encountered in solid state physics where similar mean field descriptions of the dynamics in an electron gas (e.g., the Nearly Free Electron Model) are successfully used. A rigorous justification of more general and physically more interesting situations than the one presented in this letter, however, remains to be done.

\vspace{-4pt}

\section{Acknowledgements}

We thank D. D\"urr and D.-A. Deckert for interesting discussions and helpful comments on the manuscript. Financial support of the German National Academic Foundation and the Cusanuswerk is acknowledged. 

\section{Appendix}

The variance of the potential $V$ is given by
\begin{align}
			\text {Var}_{\phi_N^{\pm f}}[V]=\mathds E_{\phi_N^{\pm f }}[V^2] -\mathds E_{\phi_N^{\pm f }}[V]^2.
\end{align}
A straightforward calculation leads to
\begin{align}
	\label{Append}
\text {Var}_{\phi_N^{\pm f}} [V] =\begin{cases} 	\sum\limits_{|j|\le \frac{N}{2}}  \frac{1}{L} \cdot \hat{v}(0)  \hspace{28,75mm} \ \ ,\ \ \phi_N^{+_1 f} \\ \sum\limits_{|j|\le \frac{N}{2}} \frac{1}{L}\cdot \Big[ \hat{v}(0)   + \sum\limits_{|k|\le \frac{N}{2}} \frac{|\hat{v}(\frac{k}{L}-\frac{l}{L}) |^2}{L}\Big],\ \ \phi_N^{+_2 f}	\\  \sum\limits_{|j|\le \frac{N}{2}} \frac{1}{L}\cdot  \sum\limits_{|k|\ge \frac{N}{2}} \frac{|\hat{v}(\frac{k}{L}-\frac{l}{L})|^2}{L} \hspace{14,4mm} ,\ \ \phi_N^{-f}	\end{cases}
\end{align}
with
\begin{align}
\hat{v}(0) =& v_0 \cdot l_0 /L\nonumber\\
\hat{v}\left(\frac{k}{L}-\frac{l}{L}\right) =& v_0 \cdot \frac{\sin(\pi l_0/L \cdot (k-l))}{(k-l)}
\end{align}
in case of the rectangular potential $v$.

\bibliographystyle{apsrev}
\bibliography{../references.bib}

\section{Remark}

The fact that there is more to say about the mean field description than calculating the average potential is, unfortunately, not always appreciated. The constancy of the average potential is, e.g., sometimes interpreted as a sufficient reason in favor of the free behavior of a single particle in a homogenous fermion gas.\\
In one of his seminal works \cite{Smoluchowski}, Marian von Smoluchowski addressed a similar fallacy while defending his microscopic view on the phenomenon of classical Brownian motion. In his answer, he gave a very clear explanation of the fact that a constant average potential does not necessarily cause a particle in a homogeneously distributed gas to move freely (the original quote can be found in \cite{Smoluchowski} on p.~762; it was translated from German by the authors):
\begin{quote}
\textit{This is the same fallacy committed by a Hazard player thinking that he could never lose an amount larger than the stake of a single dice roll. Let us investigate this analogy further. [\dots] If one takes into account, however, that the particle with mass $M$ undergoes $10^{16}$ such collisions in air, $10^{20}$ in water, most of which cancel each other with respect to the movement of the particle in $X$, but still
produce a positive or negative excess of $10^{8}$ or $10^{10}$, then one would conclude that the particle would still suffer a change in velocity of about $10^2$ or $10^4$ cm/sec.}
\end{quote}
This solved a common misunderstanding about the nature of Brownian motion at that time. It was directed against the wide spread argument that microscopic collisions can not be the cause of the erratic movement of the Brownian particle because of the fact that they disappear on average. Smoluchowski's reasoning did not depend on the particular physics of Brownian motion, i.e., a mesoscopic particle in a classical gas. The above line of thought can thus be distinguished as the original insight of the importance of statistical fluctuations which had a far reaching impact on the development of statistical physics. Here, however, Smoluchowski simply tells us what there is more to be said about the effective dynamics of a tracer particle in the homogenous gas.





\end{document}